\documentclass[12pt,preprint]{aastex}

\shortauthors{Liu et al.}

\begin{document}

\title{Determination of Solar Wind Angular Momentum and Alfv\'en Radius from Parker Solar Probe Observations}  

\author{Ying D. Liu\altaffilmark{1,2}, Chong Chen\altaffilmark{1,2}, Michael L. Stevens\altaffilmark{3}, and 
Mingzhe Liu\altaffilmark{4}} 

\altaffiltext{1}{State Key Laboratory of Space Weather, National Space 
Science Center, Chinese Academy of Sciences, Beijing, China; liuxying@swl.ac.cn}

\altaffiltext{2}{University of Chinese Academy of Sciences, Beijing, China}

\altaffiltext{3}{Smithsonian Astrophysical Observatory, Cambridge, MA 02138, USA}

\altaffiltext{4}{LESIA, Paris Observatory, PSL University, CNRS, Sorbonne University, University of Paris, France}

\begin{abstract}

As fundamental parameters of the Sun, the Alfv\'en radius and angular momentum loss determine how the solar wind changes from sub-Alfv\'enic to super-Alfv\'enic and how the Sun spins down. We present an approach to determining the solar wind angular momentum flux based on observations from Parker Solar Probe (PSP). A flux of about $0.15\times10^{30}$ dyn cm sr$^{-1}$ near the ecliptic plane and 0.7:1 partition of that flux between the particles and magnetic field are obtained by averaging data from the first four encounters within 0.3 au from the Sun. The angular momentum flux and its particle component decrease with the solar wind speed, while the flux in the field is remarkably constant. A speed dependence in the Alfv\'en radius is also observed, which suggests a ``rugged" Alfv\'en surface around the Sun. Substantial diving below the Alfv\'en surface seems plausible only for relatively slow solar wind given the orbital design of PSP. Uncertainties are evaluated based on the acceleration profiles of the same solar wind streams observed at PSP and a radially aligned spacecraft near 1 au. We illustrate that the ``angular momentum paradox" raised by R\'eville et al. can be removed by taking into account the contribution of the alpha particles. The large proton transverse velocity observed by PSP is perhaps inherent in the solar wind acceleration process, where an opposite transverse velocity is produced for the alphas with the angular momentum conserved. Preliminary analysis of some recovered alpha parameters tends to agree with the results. 

\end{abstract}


\section{Introduction}

A fundamental issue in solar and stellar physics is how the Sun sheds its angular momentum. In a landmark study, \citet[][hereinafter referred to as the WD model]{weber67} show that the angular momentum loss per unit mass can be simply expressed as $L=\Omega r_A^2$, where $\Omega$ is the solar rotation rate. The Alfv\'en radius $r_A$ is a key parameter of the Sun, at which the solar wind radial velocity changes from sub-Alfv\'enic to super-Alfv\'enic. The WD model predicts that, while there is only a modest tendency for the solar wind to co-rotate with the Sun, the magnetic stresses produce a torque to the Sun corresponding to a rigid co-rotation out to $r_A$. This process provides an efficient way for a star to spin down.

Measurements of the solar wind angular momentum flux, however, have given divergent results. Using data from Mariner 5, \citet{lazarus71} find that the magnetic field contribution to the angular momentum flux is dominated by the particle contribution. This differs from the results of \citet{pizzo83} and \citet{marsch84} based on Helios observations. \citet{pizzo83} obtain an angular momentum flux of about $0.2\times 10^{30}$ dyn cm sr$^{-1}$ near the ecliptic plane and a distribution of that flux between particles and field stresses near 1:3. They also reveal a considerable negative flux in the alpha particles, which offsets the protons' angular momentum. \citet{marsch84} suggest that the ratio between the particle and field contributions, which for average wind conditions is about 0.8, depends on the solar wind speed. Compared with the magnetic field's angular momentum content, the solar wind plasma's angular momentum flux is poorly determined. Previous results, including those from Wind measurements at 1 au \citep{finley19}, show diverse particle angular momentum fluxes; the field component of the angular momentum flux is relatively invariant. 

Launched in 2018 August, the Parker Solar Probe (PSP) mission is intended to dive below the Alfv\'en surface for the first time \citep{fox16}. PSP measurements at the first two encounters show co-rotational flows up to about 70 km s$^{-1}$ around the perihelion (35.7 solar radii), which are much larger than predicted by the WD model \citep{kasper19}. Transverse velocities exceeding WD predictions have also been reported previously. For example, observations from Mariner 5 at 0.7 au show values of about 10 km s$^{-1}$ \citep{lazarus71}. These large transverse flows challenge our understanding of the angular momentum carried by the solar wind. A specific problem concerning the PSP measurements is that the angular momentum from the high transverse velocities implies an Alfv\'en radius larger than the PSP distance from the Sun. It suggests that PSP may have already crossed the Alfv\'en surface during the first two encounters. However, the solar wind radial velocity is still super-Alfv\'enic. This discrepancy is called the ``angular momentum paradox" \citep{reville20}. 

In this Letter, we present an approach different from previous studies to determining the solar wind angular momentum flux based on PSP observations. This new approach enables us to well constrain the flux and its particle component. In the context of the approach, we illustrate that the ``angular momentum paradox" can be removed by taking into account the contribution of the alpha particles, and also make predictions on PSP crossings of the Alfv\'en surface. Improved results on the solar wind angular momentum flux may be obtained, given that (1) effects of stream-stream interactions are minimized at distances of PSP encounters; and (2) transient phenomena such as coronal mass ejections are reduced too during the present solar minimum. The outcome of this work helps clarify the angular momentum loss of the Sun as well as the origin of the large proton transverse flow.

\section{Methodology}

\subsection{Derivation of Alfv\'en Radius and Angular Momentum Flux}  

In previous studies \citep[e.g.,][]{lazarus71, pizzo83, marsch84, finley19, reville20}, the solar wind angular momentum per unit mass and the Alfv\'en radius are calculated using the expression of 
\begin{equation}
L = rv_{\phi} - {rB_rB_{\phi} \over \mu\rho v_r} = \Omega r_A^2,
\end{equation}
where $r$ is the heliocentric distance, $\rho$ the mass density of the solar wind, $B_r$ and $B_{\phi}$ the radial and azimuthal components of the magnetic field, $v_r$ and $v_{\phi}$ the corresponding velocity components, and $\mu$ the permeability constant. Contribution from the thermal anisotropy has been ignored since it is small \citep[e.g.,][]{weber70, reville20}. The angular momentum flux per steradian near the ecliptic plane is $L$ times the mass flux $\rho v_r r^2$, i.e.,         
\begin{equation}
F = \rho v_r v_\phi r^3 - {B_r B_\phi r^3 \over \mu}.
\end{equation}
The first term is the contribution from the wind plasma or the particles ($F_w$), and the second represents the magnetic field contribution ($F_m$). Indeed, use of the above formulas is straightforward, as the plasma and field parameters can all be obtained from in situ measurements. However, the first term often involves complications, such as uncertainties in the measurements of the transverse velocity and density, large fluctuations in the proton angular momentum flux arising from stream interactions and transient phenomena, and lack of accurate measurements of the alpha particles that may be a significant factor in modifying the angular momentum flux. As a result, the angular momentum flux and its distribution between the field and particles are not well constrained. 

Here we present a new approach in an attempt to avoid some of those complications. Conservation of mass ($\rho v_r r^2$) and radial magnetic flux ($r^2B_r$) leads to a constant quantity $M_A^2/(v_rr^2)$, where $M_A$ is the radial Alfv\'en Mach number defined as $M_A = v_r\sqrt{\mu\rho}/B_r$. Based on this constant quantity and $M_A=1$ at $r=r_A$, we obtain $r_A = \sqrt{v_r \over v_{rA}}{r \over M_A}$ with $v_{rA}$ being the radial velocity at the Alfv\'en critical point. The Alfv\'en radius can then be reasonably approximated as  
\begin{equation}
r_A \simeq {r \over M_A}
\end{equation}
by assuming that $v_r$ does not change much from $r_A$ to PSP encounter distances. As demonstrated by \citet{weber67}, Equation~(3) provides a rigorous estimate of the Alfv\'en radius. Once the Alfv\'en radius is obtained, the angular momentum flux can be derived using
\begin{equation}
F = \rho v_r r^2 \Omega r_A^2 \simeq {\Omega r^4 B_r^2 \over \mu v_r},
\end{equation}
which is independent of the density. A prerequisite for Equation~(4) is $L = \Omega r_A^2$, which includes both the particle and magnetic field contributions. It should be emphasized that the expression of $L = \Omega r_A^2$ is a simple, straightforward derivation from the induction equation and the azimuthal equation of motion, which hold for a variety of circumstances. As pointed out by \citet{pizzo83}, the expression remains valid even if there are substantial sources and sinks of momentum as long as they are confined to the region inside $r_A$. 

The field component of the angular momentum flux can be calculated using the second term of Equation~(2), which is independent of the particle parameters. It is relatively invariant as seen in the literature and as will be shown below. Taking advantage of the invariant nature of the field's angular momentum flux, we can then well constrain the flux carried by the wind plasma using $F_w = F - F_m$.

An advantage of the above approach is that Equation~(4) only requires the radial components of the velocity and magnetic field, which avoids various complications associated with the measurements of the transverse velocity, density, and alpha parameters. A major approximation made here is $v_r \sim v_{rA}$. We will discuss in Section~4 the error that it brings. In order to reduce errors, Equation~(4) should be used for measurements that are not far away from the Alfv\'en critical point. Also note that the expressions used here (e.g., the conservation of mass and radial magnetic flux) represent the average solar wind conditions. We expect that fluctuations, such as those from inhomogeneities, will average out, as has been done in previous studies \citep[e.g.,][]{lazarus71, pizzo83, marsch84, finley19}.

In the following analysis, we use measurements from the FIELDS instrument suite \citep{bale16} and the SWEAP package \citep{kasper16} aboard PSP during the first four encounters. The respective perihelion is 35.7 solar radii from the Sun's center for the first three encounters and 27.9 solar radii for the fourth. All the solar wind parameters are interpolated to minute averages. We cut the data at 0.3 au from the Sun. This is to reduce effects of stream-stream interactions that may develop further out, and also to diminish errors with close-in measurements.

\subsection{Effects of Alpha Particles}

There is considerable difficulty for PSP in determining the parameters of the alpha particles. SWEAP ion instruments include a Faraday Cup (SPC) that looks directly at the Sun, and an electrostatic analyzer (SPAN-I) on the ram side of PSP \citep{kasper16, case20, whittlesey20}. With SPAN-I, only part of the particle velocity distribution is observed for most of the time (D. Larson 2019, private communication). As for SPC, the alpha peak either strongly overlaps with the proton peak or is too warm to be clearly resolved against the noise for the majority of the time (M. Stevens 2020, private communication). Although our approach does not involve the alpha parameters, we want to illustrate how the alpha particles carry a significant negative angular momentum flux, which helps resolve the ``angular momentum paradox" \citep{reville20}.      

In the presence of the alpha particles, the angular momentum flux of the wind plasma can be written as
\begin{equation}
F_w \simeq n_pm_pr^3 (v_{pr}v_{p\phi} + 0.2 v_{\alpha r}v_{\alpha\phi}), 
\end{equation}
where the subscripts $p$ and $\alpha$ refer to protons and alphas respectively, $n_p$ the proton number density, and $m_p$ the proton mass. An average relative abundance of 5\% has been assumed for the alphas. We construct the alpha velocity based on Helios observations between $0.3-1$ au and data at 1 au \citep[e.g.,][]{asbridge76, marsch82}: (1) the velocity difference vector between the alphas and protons is aligned with the magnetic field and directed predominantly outward from the Sun (i.e., alphas move faster than protons); and (2) the differential speed tends to approach the local Alfv\'en speed with decreasing distance from the Sun. Given these facts, the alpha velocity in RTN coordinates at distances of PSP encounters can be approximated as     
\begin{equation}
{\bf v}_\alpha = {\bf v}_p - {\rm sign}(B_T) {{\bf B} \over \sqrt{\mu\rho}}. 
\end{equation}
The term of sign($B_T$) can be understood in the following context: since the differential streaming is aligned with the Parker spiral field, the alphas would have a negative transverse velocity (i.e., eastward). A natural thought would be whether Equations~(5) and (6) could give a particle angular momentum flux comparable to $F-F_m$ (which is the only link in the present work between the use of Equation~(4) and the analytical development on the alpha parameters). Indeed, this may serve as a test of the alpha parameter construction. However, complications certainly exist in reality. For example, the differential streaming tends to be around zero in very slow wind \citep{asbridge76, marsch82}. Therefore, Equation~(6) can only be considered as a very rough estimate. Again, we aim at illustrating if the alphas can offset a considerable amount of the angular momentum flux in the protons, not to the exact value one may expect.        

This study does not use PSP measurements of the alpha particles, which are not sufficient for a statistical sense. We do perform a preliminary analysis of some recovered alpha parameters from SPAN-I, which tends to confirm what we expect from the analytical development (see discussions in Section~4).  

\section{Observations and Results} 

Figure~1 shows the in situ measurements at the first encounter. Transverse velocities, up to about 75 km s$^{-1}$ around the perihelion, are observed. With these large co-rotational flows Equation~(1) would give Alfv\'en radii above the PSP distance from the Sun, which implies that PSP may have already crossed the Alfv\'en surface. However, the solar wind is always super-Alfv\'enic with $M_A$ larger than 3 in general. This is called the ``angular momentum paradox" \citep{reville20}. The Alfv\'en radius derived from Equation~(3) is well below the PSP distance, so the paradox can be removed. Our result from Equation~(3) is comparable to the estimate by \citet{reville20} with only the magnetic term of Equation~(1). Note that they use the SPC densities, which may be underestimated (see below).  

Negative transverse flows are also seen for enhanced radial speeds, which indicates that there might be stream-stream interactions even for distances at the encounter. The interaction between streams would result in an angular momentum transfer from the faster to slower wind, i.e., the slower wind is deflected in the direction of co-rotation while the faster wind towards the opposite; the total angular momentum, however, is conserved. The effects of stream-stream interactions are expected to increase with distance \citep[e.g.,][]{pizzo83, marsch84}.      

The proton angular momentum flux exhibits a large variance, with the standard deviation about 4.2 times the average during the time period. Negative values are seen corresponding to the eastward flows. The angular momentum flux in the magnetic field is relatively invariant with an average of about $0.08\times10^{30}$ dyn cm sr$^{-1}$, but the standard deviation is still about 3 times the average. These illustrate a common problem in the determination of the solar wind angular momentum flux and its components: the fluctuations can be much larger than the average \citep{pizzo83, marsch84}. Again, the fluctuations are expected to average out. The angular momentum flux, calculated from Equation~(4) without invoking the plasma density and the transverse velocities of the protons and alphas, shows an average value of about $0.13\times10^{30}$ dyn cm sr$^{-1}$. Application of $F_w=F-F_m$ gives an average particle flux of about $0.05\times10^{30}$ dyn cm sr$^{-1}$. An anti-correlation is visible between the angular momentum flux and the radial velocity, which can be understood from Equation~(4) if $r^2B_r$ does not vary dramatically. A similar dependence on the radial velocity is also observed in the Alfv\'en radius.

Figure~1 also indicates that, compared with the electron density from quasi-thermal noise (QTN) spectroscopy \citep{moncuquet20}, the proton density from SPC may be systematically underestimated. Since the electron density is derived from measurements of the local plasma frequency, it is thought to be most reliable. For the first four orbits the QTN density, on average, is larger than the SPC density by $30-54$\%. Although this uncertainty does not affect the determination of the angular momentum flux via Equation~(4) and the flux in the field, it may produce errors in quantities such as the proton angular momentum flux\footnote{Readers are directed to \citet{finley20} to see a different proton angular momentum flux from the SPC densities.}. We have used the QTN electron density as a proxy of the plasma density throughout the study.        

In situ measurements at the second encounter are shown in Figure~2. While the situation is in general akin to that of the first encounter, we see elevated Alfv\'en radii around the perihelion corresponding to an interval of reduced densities, with $M_A$ as low as about 1.5. This may suggest conditions in favor of PSP crossings of the Alfv\'en surface, i.e., relatively low densities and radial velocities. The third and fourth encounters are also generally similar (not shown here). For the third encounter, measurements are available only for about a third of the time; the radial Alfv\'en Mach number approaches 3 towards the perihelion. The fourth encounter yields flagged plasma data around the perihelion (data are problematic), which are removed from our study; the solar wind is still super-Alfv\'enic with $M_A$ dipping below 2 occasionally. 

Figure~3 displays the constructed alpha parameters using Equations~(5) and (6). Despite considerable fluctuations, the alpha radial velocity is generally larger than that of the protons, but similar values between the two species are also seen for slower wind. Persistent negative transverse velocities are obtained for the alphas, as expected. However, close to the perihelion the transverse velocity tends to be more positive. The alpha angular momentum flux shows an average value of about $-0.19\times10^{30}$ dyn cm sr$^{-1}$, while the proton's contribution is about $0.37\times10^{30}$ dyn cm sr$^{-1}$ on average. Although their sum is larger than $F-F_m$, the alphas are indeed capable of canceling a significant amount of the angular momentum flux carried by the protons.     

Table~1 lists the average Alfv\'en radius and angular momentum fluxes for the four encounters. Again, the angular momentum flux of magnetic stresses is relatively invariant compared with the flux in the protons. Averaging the whole data yields $r_A=9.7$ solar radii, $F=0.15\times10^{30}$ dyn cm sr$^{-1}$, and about 0.7:1 partition of the flux between the particles and field. These correspond to an average solar wind speed of 333 km s$^{-1}$. It should be stressed that the speed here should not be considered as the final speed, as the solar wind may be still accelerating at distances of the encounters. We see both similarities and differences, comparing our results with previous ones \citep{lazarus71, pizzo83, marsch84, finley19} as mentioned in Section~1. In particular, our flux is comparable to the lower end of the range obtained by \citet{pizzo83}, and the ratio between the particle and field contributions is similar to that given by \citet{marsch84}. The alpha particles, in principle, could cancel a considerable amount of the flux in the protons. However, we find this more difficult as we move closer to the Sun. It is because either not enough flow states are sampled by the spacecraft, or the alpha velocity deviates from Equation~(6). On average, the alphas are anticipated to carry an angular momentum flux of about $-0.5\times10^{30}$ dyn cm sr$^{-1}$ in order to bring down the protons' flux to the level of the wind plasma.    

Table~2 gives the Alfv\'en radius and angular momentum fluxes as a function of the solar wind speed. The field component of the flux is remarkably constant over the speed. The total flux decreases with speeds, and so does the flux in the wind plasma given the invariance of the flux in the field. We see a predominance of the plasma contribution over the field's for speeds less than 250 km s$^{-1}$, a near equipartition for speeds between 250 and 350 km s$^{-1}$, and a reversal for speeds larger than 350 km s$^{-1}$. Interestingly, the wind plasma carries zero or even negative angular momentum flux when the speed exceeds 450 km s$^{-1}$; we expect a negative flux of about $0.5\times10^{30}$ dyn cm sr$^{-1}$ in the alphas for such a complete cancellation. Predictions can also be made on PSP Alfv\'en surface crossings based on the speed dependence of the Alfv\'en radius. The closest approach to the center of the Sun is 9.86 solar radii and will occur during the final three orbits \citep{fox16}. Substantial diving below the Alfv\'en surface is plausible only for the solar wind with speeds below 350 km s$^{-1}$. For speeds above 350 km s$^{-1}$ the chance seems low, although PSP may scratch the Alfv\'en surface under some circumstances.  

\section{Conclusions and Discussions}

The approach that we have presented does not require the transverse velocities of the solar wind protons and alphas, thus avoiding various complications. This may be helpful for PSP measurements of the Alfv\'en radius and angular momentum flux, which exhibit large proton transverse flows of unknown origin and difficulty in determining the alpha parameters. With the approach the angular momentum flux and its distribution between the magnetic field and particles are well constrained. Averaging data from the first four encounters within 0.3 au gives an Alfv\'en radius of 9.7 solar radii, a flux of $0.15\times10^{30}$ dyn cm sr$^{-1}$ near the ecliptic plane, and a partition 0.7:1 of that flux between the particles and field. The angular momentum flux in the field is remarkably constant over the solar wind speed, whereas the total flux and thus its particle component decrease with speeds. We expect zero or even negative angular momentum flux in the particles when the speed exceeds 450 km s$^{-1}$. A speed dependence in the Alfv\'en radius is also observed, so we anticipate a ``rugged" Alfv\'en surface around the Sun. PSP's persistent diving below the Alfv\'en surface seems plausible only for the solar wind with speeds below 350 km s$^{-1}$. 

In the context of our approach the ``angular momentum paradox" \citep{reville20} is removed. By constructing the alpha velocity based on observations from $0.3-1$ au, we illustrate that, in principle, the alpha particles can cancel a considerable amount of the angular momentum flux in the protons. On average a flux of about $-0.5\times10^{30}$ dyn cm sr$^{-1}$ in the alphas is needed to reduce the protons' flux to the level of the wind plasma. A recent work by \citet{fisk20} suggests that the large proton transverse velocity is a result of interchange reconnection in the corona. If this is true, we would expect both positive and negative transverse flows everywhere. However, persistent positive flows are observed around all the perihelions. Our experiment on the alphas indicates that the large transverse flow is likely inherent in the solar wind acceleration process. The acceleration process is such that a positive transverse velocity is produced for the protons while a negative one for the alphas, but the angular momentum is conserved. 

Recovery of the alpha parameters from PSP observations will be key to testing the results here and understanding the origin of the large proton transverse flows. Fortunately, around the perihelion of the fourth encounter PSP moves fast enough to shift the majority of the alpha core into the field of view of SPAN-I. Our preliminary analysis of the recovered alpha parameters yields results similar to what we have expected: for the slower wind, the alpha radial velocity resembles that of the protons, and the alpha transverse velocity fluctuates around zero; for the faster wind, the alphas show a radial velocity larger than that of the protons and a significant negative transverse velocity. These results are consistent with a considerable negative angular momentum flux in the alphas and our suggestion on the origin of the large proton transverse flows. We note a recent work by \citet{finley21} looking at the alpha particles from SPAN-I measurements, which is coincident with the time frame of the present study. They perform bi-Maxwellian fits of the truncated velocity distribution functions of the particles from the third and fourth encounters. Their results roughly agree with our preliminary analysis of the recovered alpha parameters, in terms of the velocity difference between the protons and alphas for both the slower and faster wind. However, the data are limited and not sufficient for a statistical analysis. Studies better tackling calibration issues are needed for a definite conclusion.       

Note that Equation~(3) gives a lower limit of the Alfv\'en radius, although it should be close to the real value \citep{weber67}. Detailed error analysis is difficult without knowing the velocity profile as a function of distance. For reference, if the radial velocity grows by 10\% from the Alfv\'en critical point to PSP encounter distances within 0.3 au, the value of the Alfv\'en radius will increase by about 5\% (i.e., $\sqrt{v_r \over v_{rA}}$), and the angular momentum flux will increase by 10\%. We have identified the same solar wind streams observed at PSP and a spacecraft near 1 au when they are radially aligned. The results show quite different acceleration profiles for different solar wind streams. As an extreme case (2020 January 28 to 31), the radial velocity increases from about 238 km s$^{-1}$ at 29 solar radii to about 512 km s$^{-1}$ at Wind, implying an average acceleration of about 1.47 km s$^{-1}$ per solar radius. Another case on 2019 April 7 to 12 indicates an average acceleration of only about 0.39 km s$^{-1}$ per solar radius, i.e., 363 km s$^{-1}$ at 39 solar radii versus 432 km s$^{-1}$ at Wind. If the acceleration is assumed to be constant from the Alfv\'en critical point ($\sim$10 solar radii) to 1 au, these two cases suggest an increment in the radial velocity by about 13\% and 3\%, respectively, from the Alfv\'en critical point to the PSP distances.

Another issue is that PSP may not have collected enough flow states. We await more data in the years to come, with improved calibration and in particular considerable recovery of the alpha parameters.   

\acknowledgments The research was supported by NSFC under grant 41774179, Beijing Municipal Science and Technology Commission (Z191100004319003), and the Specialized Research Fund for State Key Laboratories of China. We acknowledge the NASA Parker Solar Probe mission and the SWEAP and FIELDS teams for use of data.

\clearpage

\begin{figure}
\epsscale{0.9} \plotone{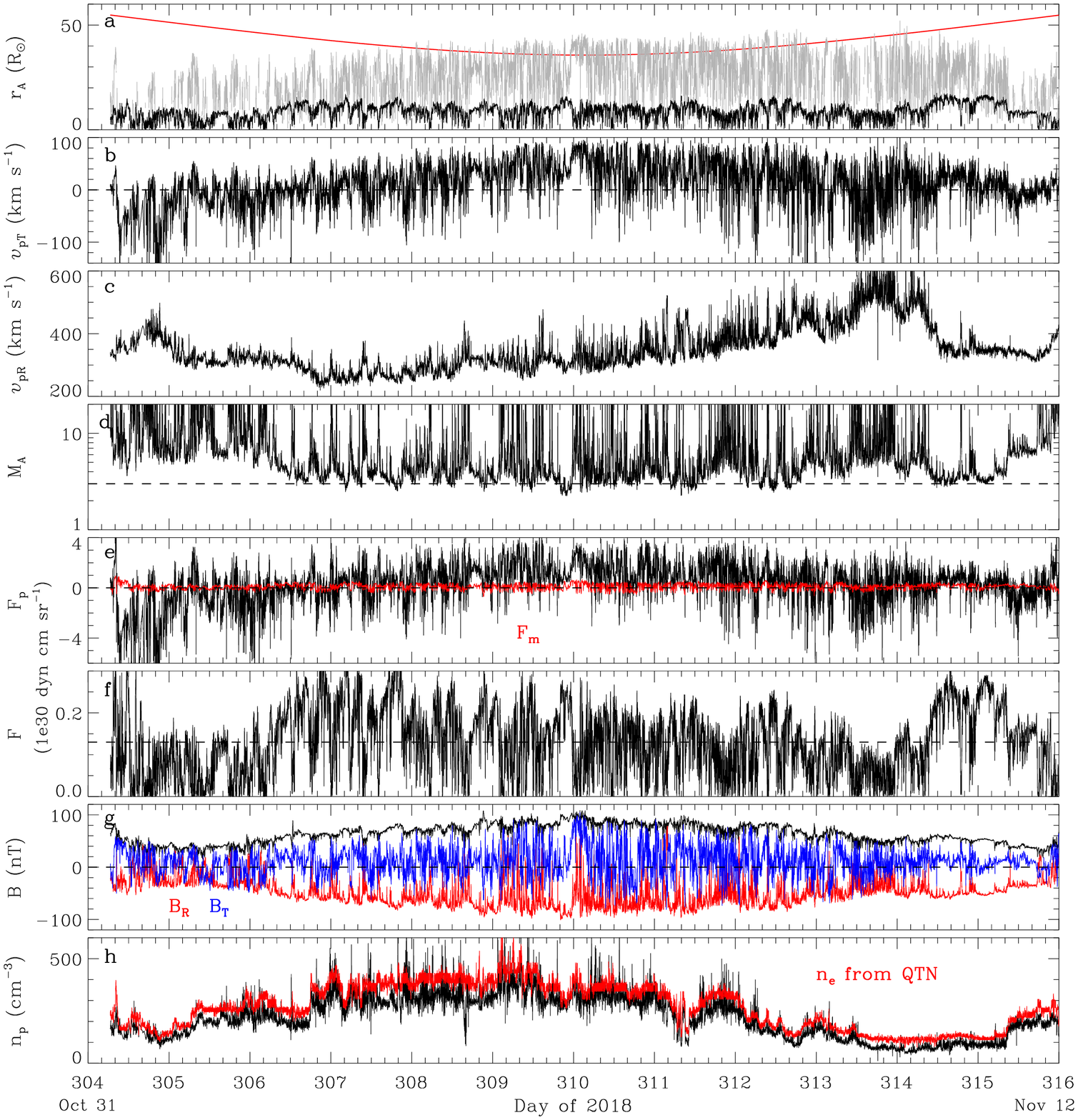} 
\caption{PSP measurements at the first encounter. (a) Alfv\'en radius in comparison with the distance of the spacecraft (red). The gray curve represents the Alfv\'en radius derived from Equation~(1). (b)-(c) Proton transverse and radial velocities. (d) Radial Alfv\'en Mach number. (e) Proton angular momentum flux in comparison with the angular momentum flux of magnetic stresses (red). (f) Angular momentum flux calculated from Equation~(4). (g) Magnetic field strength and radial (red) and transverse (blue) components. (h) Proton density in comparison with the electron density (red) from quasi-thermal noise (QTN).}  
\end{figure}

\clearpage

\begin{figure}
\epsscale{0.9} \plotone{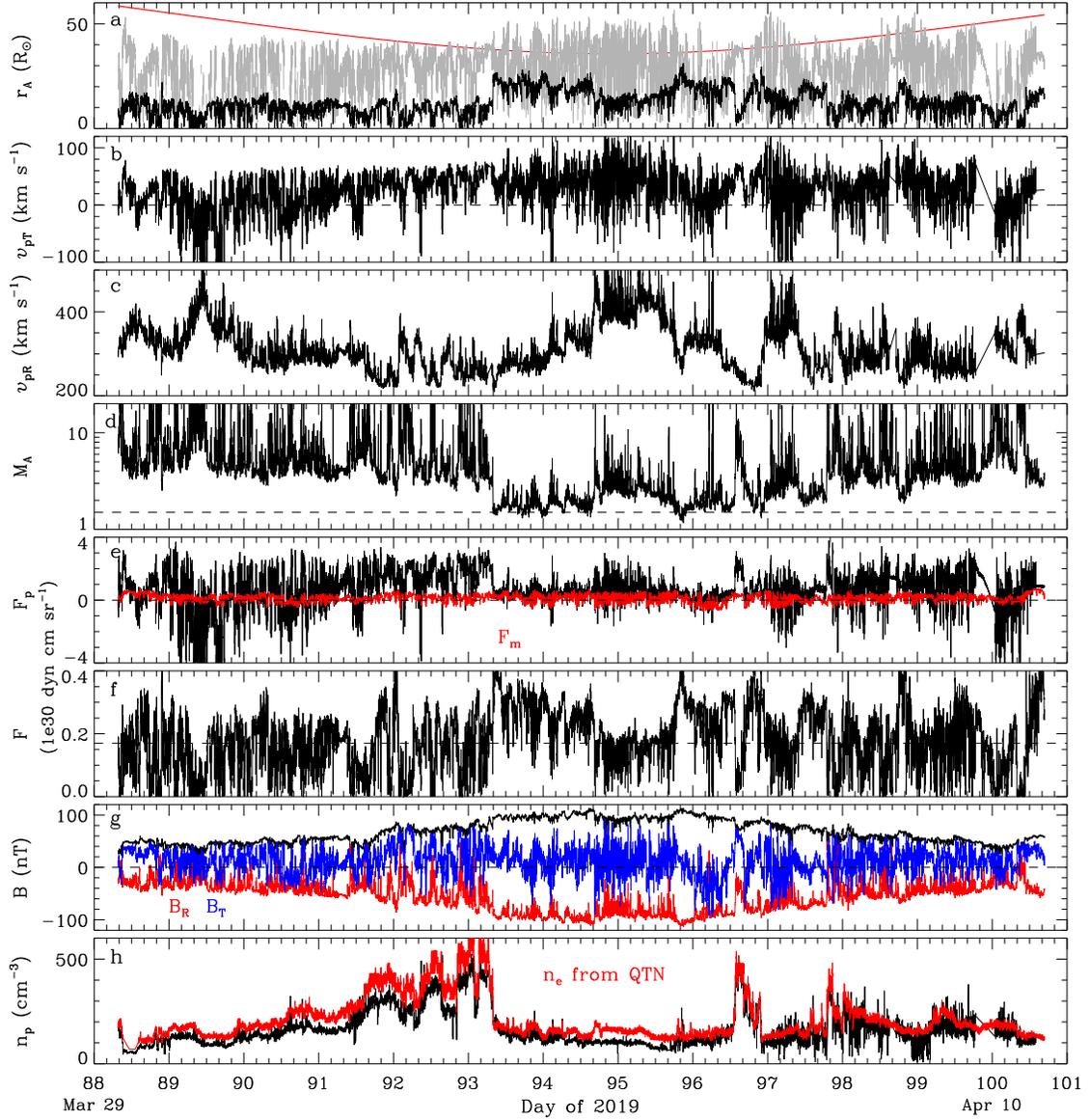} 
\caption{PSP measurements at the second encounter. Similar to Figure~1. The radial Alfv\'en Mach number drops to about 1.5 around the perihelion.}
\end{figure}

\clearpage

\begin{figure}
\epsscale{0.9} \plotone{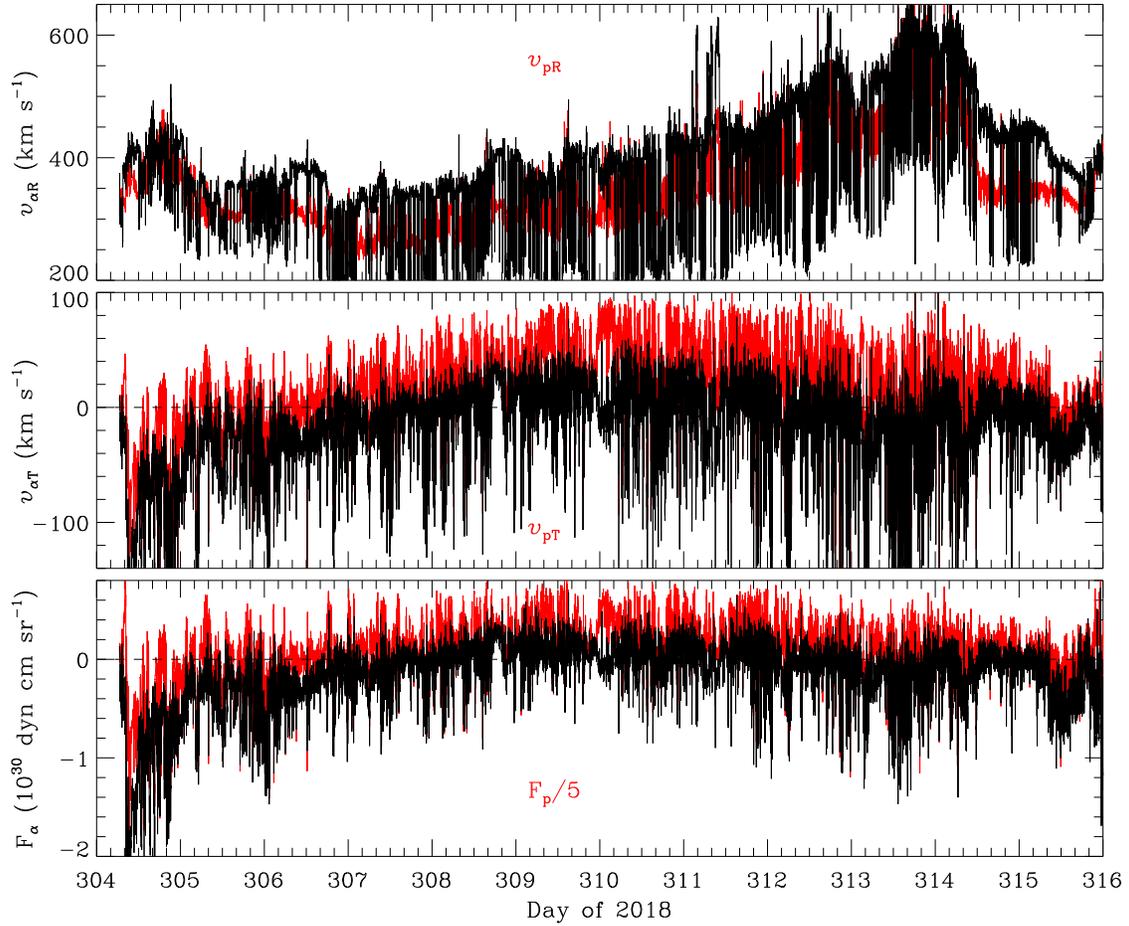} 
\caption{Constructed helium-ion velocity components and angular momentum flux in comparison with the proton counterparts (red) for the first encounter. The proton angular momentum flux is divided by a factor of 5.}
\end{figure}

\clearpage

\begin{deluxetable}{ccccccc}
\tablecaption{PSP Measurements of Alfv\'en Radius and Angular Momentum Flux} 
\tablewidth{0pt}
\tablehead{
\colhead{Encounter} & \colhead{$r_A$} & \colhead{$F$\tablenotemark{a}} & \colhead{$F_m$}  
& \colhead{$F_w$\tablenotemark{b}} & \colhead{$F_p$} & \colhead{$F_{\alpha}$\tablenotemark{c}} \\
 & ($R_{\odot}$) & \multicolumn{5}{c}{($10^{30}$ dyn cm sr$^{-1}$)} }
\startdata
1            & 8.1   & 0.13 & 0.08 & 0.05 & 0.37 & $-0.19$ \\
2            & 11.8 & 0.17 & 0.13 & 0.04 & 0.66 & $-0.08$ \\
3            & 9.7  & 0.15 & 0.07 & 0.08 & 0.47 & $-0.13$ \\
4            & 9.0  & 0.13 & 0.08 & 0.06 & 0.82 & $-0.01$ \\
Average & 9.7  & 0.15 & 0.09 & 0.06 & 0.58 & $-0.10$ \\
\enddata
\tablenotetext{a}{Angular momentum flux calculated from Equation~(4).}
\tablenotetext{b}{Angular momentum flux of the solar wind plasma calculated from $F_w=F-F_m$.} 
\tablenotetext{c}{Constructed helium-ion angular momentum flux based on observations from $0.3-1$ au.}
\end{deluxetable}

\clearpage

\begin{deluxetable}{ccccccc}
\tablecaption{Alfv\'en Radius and Angular Momentum Flux as a Function of Speed}  
\tablewidth{0pt}
\tablehead{
\colhead{Speed\tablenotemark{a}} & \colhead{Percentage} & \colhead{$r_A$} & \colhead{$F$} 
& \colhead{$F_m$} & \colhead{$F_w$\tablenotemark{b}} & \colhead{$F_p$} \\
(km s$^{-1}$) & & ($R_{\odot}$) & \multicolumn{4}{c}{($10^{30}$ dyn cm sr$^{-1}$)} } 
\startdata
$v_p\leqslant 250$            & 4.5\%   & 12.3 & 0.22 & 0.08 & 0.14 & 0.86 \\
$250 < v_p\leqslant 350$  & 62.3\% & 10.2 & 0.16 & 0.09 & 0.07 & 0.65 \\ 
$350 < v_p\leqslant 450$  & 27.8\% & 9.1   & 0.12 & 0.09 & 0.03 & 0.42 \\
$v_p> 450$                       & 5.4\%   & 7.9   & 0.09 & 0.10 & -0.01 & 0.50 \\
\enddata
\tablenotetext{a}{The solar wind at distances of PSP encounters may be still accelerating, so the speed here should not be confused with the speed at 1 au.}
\tablenotetext{b}{Angular momentum flux of the solar wind plasma calculated from $F_w=F-F_m$.}
\end{deluxetable}

\end{document}